\documentclass[twocolumn,showpacs,amsmath,floatfix,prl]{revtex4} 
\usepackage{graphicx}
\begin{document}

\title{Optical Response of Solid CO$_2$ as a Tool for the Determination of the High Pressure Phase}

\author{S. Sharma}
\email{sangeeta.sharma@uni-graz.at}
\author{J. K. Dewhurst}
\author{C. Ambrosch-Draxl}
\affiliation{Institute for Theoretical Physics, Karl--Franzens--Universit\"at Graz,
Universit\"atsplatz 5, A--8010 Graz, Austria.}

\date{\today}

\begin{abstract}
We report first-principles calculations of the frequency dependent linear and second-order optical properties of the two probable extended-solid phases of 
CO$_2$--V, i.e.  $I\overline42d$ and $P2_12_12_1$. Compared to the parent $Cmca$ phase the linear optical susceptibility of both phases is much smaller. 
We find that $I\overline42d$ and $P2_12_12_1$ differ substantially in their linear optical response in the higher energy regime. The nonlinear optical 
responses of the two possible crystal structures differ by roughly a factor of five. 
Since the differences in the nonlinear optical spectra are pronounced in the low energy regime, i.e. below the band gap of diamond, measurements 
with the sample inside the diamond anvil cell are feasible.  We therefore suggest optical experiments 
in comparison with our calculated data as a tool for the unambiguous identification of the high pressure phase of CO$_2$.
\end{abstract}
        
\maketitle

A few years after the synthesis of the extended-solid phase of CO$_2$ at 40 GPa the crystal structure of this phase is still not known.  
This phase was found to be similar to one of the noncentrosymmetric polymorphs of quartz \cite{iota99,serra99} and was designated as CO$_2$--V.
Since solid CO$_2$ has very high tensile strength (comparable to that of boron nitride \cite{yooc99}) this discovery could be very useful in device applications.
There is also considerable interest in the role of CO$_2$ in planetary geology \cite{hoffman01}.

Experimental work done to identify the exact crystal structure of CO$_2$ at high pressure suggests a low symmetry tridymite phase ($P2_12_12_1$) \cite{yoo99}. 
The presence of a nonlinear optical signal  also indicated the absence of inversion symmetry in the crystal structure \cite{iota99}. Most of the theoretical 
work has been concentrated around the lattice relaxations and total energy calculations in search of the minimum energy structure \cite{dong00I,holm00,dong00II}.
Theory predicts the $\beta$-cristobalite phase ($I\overline42d$) to be lower in energy than $P2_12_12_1$ by around 0.19 eV  per formula unit \cite{dong00II}.
However, none of the SiO$_2$-like polymorphs of CO$_2$ gave a satisfactory fit to the experimental XRD data \cite{yoo99,holm00}. 
Other possible quartz polymorphs have been eliminated by their much higher total energies \cite{dong00I}.
Since all experiments were performed at elevated temperatures (1800K) and the theoretical energy difference between the two phases is small, the energy 
minimization is not sufficient to solve the discrepancy between theory and experiments and to predict the actual high pressure phase. In order to understand 
the nature of this phase transition there is the need for analyzing other physical properties and finding a satisfactory agreement between theory and experiment.
 Since the two probable phases have very different electronic structure it is possible to distinguish them by their optical response. So far, 
neither a theoretical or experimental study of the linear optical response have been performed. Although second harmonic generation (SHG) by CO$_2$ has been
 seen experimentally, there exist no experimental data or calculations of the frequency dependent SHG susceptibility.  
 The aim of the present work is to show that the linear and nonlinear optical spectra indeed allow an unambiguous identification of the CO$_2$--V crystal 
structure. To this extent we have calculated the linear and second-order optical response of solid carbon dioxide, CO$_2$--V, in the two possible phases. 
The linear optical spectra of these probable phases are compared to each other and also to that of the $Cmca$ phase, which under high pressure and temperature 
conditions 
leads to CO$_2$--V.  Since quartz is often used as a standard in second-harmonic (SH) experiments, a detailed comparison is made between the second-order 
optical response of the two CO$_2$--V phases and that of SiO$_2$ ($I\overline42d$ phase). We propose that on basis of our results, future experimental work 
will help to resolve the present disagreement between theory and experiment and hence clarify the high pressure phase of CO$_2$. 

Total energy calculations are performed using the state-of-the-art full-potential linearized augmented plane wave FPLAPW method implemented in the {\sf WIEN2k} 
code \cite{WIEN}. For the exchange-correlation potential we use the generalized gradient approximation (GGA) derived by Perdew and Wang \cite{perdew92}. 
The  detailed formalism for the determination of the linear susceptibility tensor $\chi^{(1)}(\omega)$ within the FPLAPW formalism has been presented
 before \cite{cad}. The susceptibility for the SHG $\chi^{(2)}(2\omega,\omega,\omega)$ has been calculated using an extension to this program \cite{sl}. 
The SHG susceptibility obtained satisfies all the theoretical sum rules presented by Scandolo and Bassani \cite{scandolo95}.
All the calculations are converged in terms of basis functions as well as in the size of the $k$-point mesh representing the Brillouin zone. 
The optical properties are calculated on a mesh of 1500 $k$-points in the irreducible wedge of the Brillouin zone.

The phases of CO$_2$ under investigation in the present work are the $Cmca$ phase and the two probable extended-solid phases for CO$_2$--V,
namely $I\overline42d$ and $P2_12_12_1$.  For the $P2_12_12_1$ phase the starting point is the experimental lattice parameters \cite{yoo99}
with the atomic positions of the $C222_1$ phase (which is the supergroup of the space group $P2_12_12_1$). In a first step, the lattice parameters are 
kept constant at the experimental values and the internal parameters are fully relaxed. This structure is referred
to as $P2_12_12_1$--${expt}$ in the present work. Next, the lattice and the internal parameters are simultaneously relaxed (keeping a
constant volume \cite{remark_vol}) to attain the minimum energy configuration.  This structure is referred to as
$P2_12_12_1$--${rel}$.  The energy lowering on going from $P2_12_12_1$--${expt}$ to the $P2_12_12_1$--${rel}$ structure is 0.81
eV per formula unit, which is around 0.16 eV more than the previous theoretical value \cite{dong00II}. The larger energy in the present calculations compared 
to the previous theoretical value could be due to the difference in the method of calculation, since in the previous work the pseudopotential approach was 
used \cite{dong00II}.  In case of the $I\overline42d$ phase, due to the lack of experimental data, the starting lattice constants and internal parameters
are taken to be those of SiO$_2$ in the same phase.  Then the atoms as well as the cell volume are fully relaxed to attain the minimum energy configuration.
The lattice parameters for all the structures are presented in Table I.
In accordance with the previous theoretical work \cite{holm00,dong00II} we find $\beta$-cristobalite ($I\overline42d$) to be lower in energy compared to 
the $P2_12_12_1$--${rel}$ phase.  The difference in energy between the $P2_12_12_1$--${rel}$ and  the theoretically determined 
$I\overline42d$ phase at 40 GPa is 0.38 eV/CO$_2$.

\begin{figure}[ht]
\centerline{\includegraphics[scale=0.5,width=\columnwidth,angle=-90]{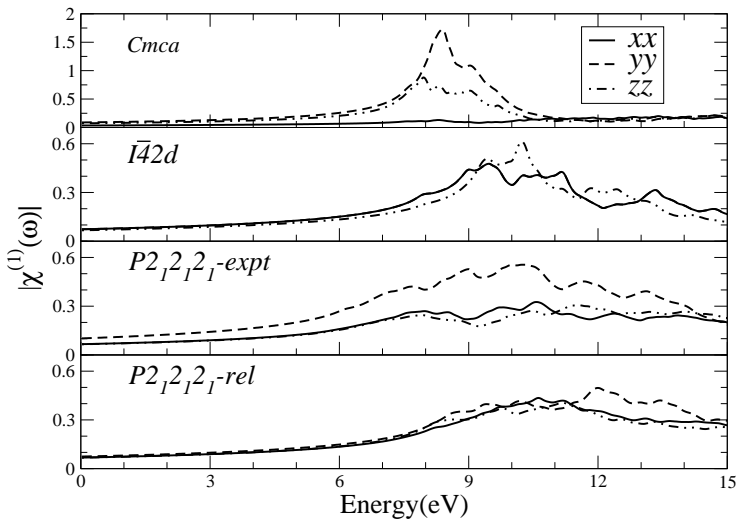}}
\caption {The magnitude of the linear susceptibility components for solid CO$_2$ in 
(a) the $Cmca$ phase, 
(b) the $I\overline42d$ phase at 40 GPa, 
(c) the $P2_12_12_1$--$expt$ structure and 
(d) the $P2_12_12_1$--$rel$ structure. }
\label{fig_LO}
\end{figure}
\begin{table}[h]
\caption {Lattice parameters for various phases. The cell parameters are in \AA. The atomic positions are in lattice coordinates.}
\centerline{
\begin{tabular}{|c| c c c| c c | } \hline
Phase                   & $a$ (\AA) & $b$ (\AA) & $c$ (\AA) & atom   & position (a,b,c)    \\ \hline
$I\overline42d$         & 4.162     & 4.162     &  5.886    & C      & (0,0,0)             \\ 
                        &           &           &           & O      & (0.880,0.250,0.125) \\ \hline
$P2_12_12_1$--${expt}$  & 6.214     & 4.351     &  6.064    & C(1)   & (0.434,0.620,0.690) \\ 
                        &           &           &           & C(2)   & (0.920,0.940,0.710) \\
                        &           &           &           & O(1)   & (0.890,0.599,0.200) \\ 
                        &           &           &           & O(2)   & (0.550,0.200,0.020) \\ 
                        &           &           &           & O(3)   & (0.228,0.577,0.214) \\ 
                        &           &           &           & O(4)   & (0.469,0.257,0.351) \\ \hline
$P2_12_12_1$--${rel}$   & 6.767     & 3.773     & 6.420     & C(1)   & (0.900,0.920,0.700) \\ 
                        &           &           &           & C(2)   & (0.420,0.494,0.690) \\
                        &           &           &           & O(1)   & (0.589,0.190,0.990) \\ 
                        &           &           &           & O(2)   & (0.247,0.703,0.211) \\ 
                        &           &           &           & O(3)   & (0.430,0.220,0.293) \\ 
                        &           &           &           & O(4)   & (0.934,0.757,0.161) \\ \hline
\end{tabular}
}
\end{table}

The results for the linear optical spectra of CO$_2$ in various phases are presented in Fig. \ref{fig_LO}. None of the optical spectra are scissors corrected.
 Both the $P2_12_12_1$ and the $I\overline42d$ phase are taken at 40 GPa. The $Cmca$ phase shows a strong anisotropy with 
$|\chi^{(1)}_{yy}(\omega)|$ much larger in amplitude than  $|\chi^{(1)}_{xx}(\omega)|$, and $|\chi^{(1)}_{zz}(\omega)|$ lying in between the two. The reason for
 this anisotropy is the orientation of the molecules within the crystal.  The neighbouring molecules are parallel to each other in the $ab$ plane and form 
a herringbone pattern in the $ac$ and $bc$ planes [Fig. 2(b) in Ref. \onlinecite{yoo02}]. This causes the $yy$ and $zz$ polarized transition matrix elements 
(TMEs) to be large.  On the other hand the $xx$ polarized TMEs are close to zero, leading to small values for $|\chi^{(1)}_{xx}(\omega)|$.  
$|\chi^{(1)}_{zz}(\omega)|$ is smaller than $|\chi^{(1)}_{yy}(\omega)|$ because the lattice parameter $b$ is smaller than $c$. 
On going from the low pressure $Cmca$ phase to any other phase of CO$_2$-V, $|\chi^{(1)}_{yy}(\omega)|$ and the anisotropy between 
$|\chi^{(1)}_{yy}(\omega)|$ and $|\chi^{(1)}_{xx}(\omega)|$ decrease dramatically.
The latter fact can again be explained by the orientation of the molecules inside the unit cell  [Fig. 3E and 3F in Ref. \onlinecite{holm00}]. 
The neighbouring CO$_2$ molecules are no more parallel or perpendicular in any direction leading to a weakening of the very strong transitions and an
enhancement of the TMEs which were forbidden in the $Cmca$ phase.  
For all the probable phases of CO$_2$--V the spectral weights are shifted to higher energies. 
The optical band gap of the $P2_12_12_1$--${expt}$  structure 
is smaller than that of the $Cmca$ phase, while the other two structures show almost equal optical band gaps, which are higher than that of the $Cmca$ phase. 
Compared to the $Cmca$ phase the maximum peak height is around 3 times smaller in the $I\overline42d$ phase, while it is about 4 times 
smaller in the $P2_12_12_1$ structures. Beyond 11 eV, the response of all the probable phases of CO$_2$--V is on average similar in magnitude, but differs in 
the amount of anisotropy. 

For  $I\overline42d$ there are two independent tensor components. Compared to the $zz$ component, the $xx$ component is approximately upto 2 times as large in 
the energy range below 10 eV. Upto 10.5 eV and between 11.5 and 13 eV the effect is opposite, while the two are nearly the same for higher energies.
For the  $P2_12_12_1$ structures there are three inequivalent components and the anisotropy between the $xx$ and $zz$ components is very small
compared to the $I\overline42d$ phase.  We also find the lattice relaxation effects  which have already been described in
detail by Dong {\it et al.} \cite{dong00II}. They have some consequences for the optical spectra. 
On going from the $Cmca$ phase to $P2_12_12_1$--${expt}$, 
the optical band gap decreases, while it increases in case of $P2_12_12_1$--${rel}$. Another noticeable feature is that the anisotropy is much larger 
in the $P2_12_12_1$--${expt}$ structure. 
Since CO$_2$--V is not quenchable into ambient pressure at room temperature the optical measurements need to be performed with the
sample inside the diamond anvil cell (DAC). The band gap of diamond is 5.5eV. Therefore the energy regime below 5.5eV is most interesting, since the 
optical response in this energy regime is purely from the sample inside the DAC.
The difference between the linear optical spectra of various
probable phases of CO$_2$ below this energy range is not as pronounced as in the higher energy regime. The results though are of theoretical interest
and help in identification of various features in the nonlinear optical spectra (which is rather difficult to do from the band structure plots), but may not 
be very good in phase distinction under these experimental conditions.
\begin{figure}[ht]
\centerline{\includegraphics[width=\columnwidth,angle=-90]{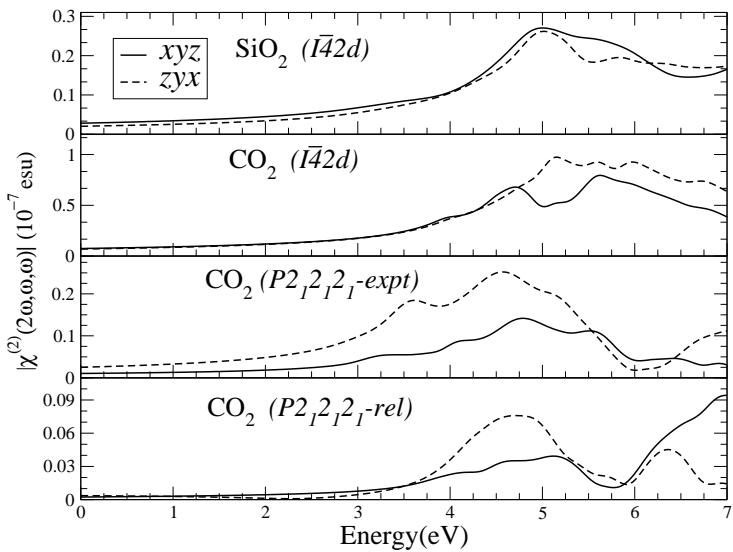}}
\caption {The magnitude of the second-harmonic generation susceptibility components $xyz$ and $zyx$ for
(a) SiO$_2$ in the $I\overline42d$ phase, 
(b) CO$_2$ in the $I\overline42d$ phase at 40 GPa, 
(c) in the $P2_12_12_1$--$expt$ structure and 
(d) in the $P2_12_12_1$--$rel$ structure.}
\label{fig:NLO}
\end{figure}

The nonlinear optical properties are much more sensitive to small changes in the band structure than the linear optical spectra. The calculated
magnitude of the SHG susceptibility by CO$_2$--V in various probable phases are presented in Fig. \ref{fig:NLO}.  Due to the presence of inversion symmetry 
in the $Cmca$ phase the SHG susceptibility is zero. The second-order response of SiO$_2$ ($I\overline42d$), which is often used as standard in SHG experiments, 
is also presented in the figure. In the energy range between 3 and 5.5 eV the magnitude of the SHG susceptibility by CO$_2$ in the $P2_12_12_1$--${expt}$ phase 
is 1.5--2 times smaller than that of quartz. On the other hand the $I\overline42d$ phase of CO$_2$ shows a SHG susceptibility 3--4 times larger than that of 
SiO$_2$. These pronounced differences make the phases easily identifiable. The energy range of 3--5.5 eV lies below the band gap of diamond and corresponds 
to laser light of wavelengths between 413.3 - 247.9 nm. Previous SH experiments were performed using laser light of 1054 nm wavelength \cite{iota99}, 
which correspond to 1.176 eV. At this energy our calculated value of $|\chi_1^{(2)}(2\omega, \omega, \omega)|$ for CO$_2$ in the $I\overline42d$ phase is
2.7 times larger than that of SiO$_2$. While, the SHG susceptibility by CO$_2$ in the $P2_12_12_1$--${expt}$ phase is 2.6 times smaller than that of SiO$_2$.  
The previous experimental work \cite{yoo99} reports the SH conversion efficiency of the CO$_2$--V phase, which is not just proportional to the square of the SHG 
susceptibility but also depends on the input intensity and sample parameters. So no direct comparison of the existing experimental data can be made with 
the present results. Since the difference in the second-order response by the probable phases of CO$_2$-V are more pronounced in the energy range between 
3 and 5.5 eV than in between 1--2 eV, experiments at higher energies would be favourable. Experimental determination of the SHG susceptibility by CO$_2$--V 
and SiO$_2$, over some frequency range, and a comparison with our work could lead to conclusive results. 

The two structures of the $P2_12_12_1$ phase are also differentiable by their SHG response. Due to the large optical band gap of  $P2_12_12_1$--${rel}$ 
the SHG susceptibility is approximately 3 times smaller than that of  $P2_12_12_1$--${expt}$.
As shown in Ref. \onlinecite{dong00II} the XRD patterns for neither of these structures matches the experiments. The theoretically calculated 
XRD pattern of $P2_12_12_1$--$rel$ shows a better agreement with the experimental XRD pattern at higher values of 2$\theta$ (9--13 degrees), while the 
XRD pattern for $P2_12_12_1$--$expt$ is in better agreement at low values of 2$\theta$ (6--8 degrees). It is possible that under the experimental conditions 
of high temperatures the material is a polycrystalline mixture of these two structures or has a configuration in between the two, in which case the difference 
between the optical spectra of the two phases would  represent a measure of the uncertainty in the optical response of this phase. 
Despite this fact, we want to point out that in general the differences in the optical response between the two phases are much more pronounced than 
``the lattice relaxation effects'' for the $P2_12_12_1$ phase making them distinguishable by optical probes.

The optical coefficients in the static limit are directly measurable quantities. The calculated inequivalent components of $\epsilon_1(0)$ and 
$|\chi^{(2)}(0)|$ are presented in the Table II.

\begin{table}[h]
\caption {The static dielectric constants $\epsilon^{xx}_1(0)$, $\epsilon_1^{yy}(0)$ and $\epsilon^{zz}_1(0)$, and the SHG susceptibility components 
in the static limit $|\chi^{(2)}_{zyx}(0)|$ and $|\chi^{(2)}_{xyz}(0)|$ in $\times 10^{-8}$ esu for CO$_2$ in various phases.}
\centerline{
\begin{tabular}{|c| c c c| c c | } \hline
Phase                   &  $\epsilon_1^{xx}(0)$ & $\epsilon_1^{yy}(0)$ & $\epsilon_1^{zz}(0)$ & $|\chi^{(2)}_{xyz}(0)|$ & $|\chi^{(2)}_{zyx}(0)|$ \\ \hline
$Cmca$                  &  1.87                 & 3.23                   & 2.84               & 0.00              & 0.00              \\ \hline
$I\overline42d$         &  2.82                 & 2.82                   & 2.66               & 0.74              & 0.69              \\ \hline
$P2_12_12_1$--${expt}$  &  2.69                 & 3.57                   & 2.67               & 0.10              & 0.25              \\ \hline
$P2_12_12_1$--${rel}$   &  2.69                 & 2.86                   & 2.68               & 0.02              & 0.04              \\ \hline
\end{tabular}
}
\end{table}

In the $Cmca$ phase, the anisotropy between $\epsilon_1^{xx}(0)$ and $\epsilon_1^{yy}(0)$ is 54\%, and between $\epsilon_1^{xx}(0)$ and $\epsilon_1^{zz}(0)$ 
it is 41\%. In the $I\overline42d$ phase, the anisotropy between the $xx$ and $zz$ components is 6\% only. For the tridymite phase at experimental lattice 
parameters ($P2_12_12_1$--${expt}$)  the maximum anisotropy found between $\epsilon_1^{yy}(0)$ and $\epsilon_1^{zz}(0)$ (28 \%), while between 
$\epsilon_1^{xx}(0)$ and $\epsilon_1^{zz}(0)$ it is only 1\%.  In the case of $P2_12_12_1$--${rel}$ the anisotropy between the $xx$ and $yy$ 
components is 7\%,  and between the $xx$ and $zz$ components is close to zero. 
 
Amongst the phases under investigation in the present work the anisotropy in the $xyz$ and $zyx$ components of $|\chi^{(2)}(0)|$ is  maximum for 
$P2_12_12_1$--${expt}$ with 86 \% and minimum for $I\overline42d$ with 7\%. $P2_12_12_1$--${rel}$  lies in between the two with an anisotropy of 67 \%.
For SiO$_2$ in the $I\overline42d$ phase, our calculated values of $|\chi_{xyz}^{(2)}(0)|$ and $|\chi_{zyx}^{(2)}(0)|$ are 0.28 and 0.20 $\times 10^{-8}$ esu, 
respectively. It may be noted that the anisotropy as well as the magnitude of $\chi_1^{(2)}(0)$ for CO$_2$ in the $P2_12_12_1$--${expt}$ phase are close to 
that of SiO$_2$. The fact that $|\chi^{(2)}(0)|$ of CO$_2$ is around 3 times larger in the $I\overline42d$ phase, and 5--10 times smaller in the 
 $P2_12_12_1$--${rel}$ phase compared to SiO$_2$ makes the distinction between the two phases clear and hence the identification 
of the high pressure phase feasible.

In summary we find that the optical properties are excellent tools for an unambiguous determination of the high pressure phase of CO$_2$.
Particularly, the second-order response of CO$_2$--V compared to that of SiO$_2$ can be used for phase determination even with the sample inside the DAC.
Experiments of this kind and a detailed comparison 
with the present theoretical results would shed light on the existing discrepancy between 
theory and experiments regarding the space group of the CO$_2$--V phase.
 
The work is supported by the Austrian Science Fund (projects P13430, P14004 and P16227) and the EXCITING network funded by the EU (Contract HPRN-CT-2002-00317).
 One of the authors (SS) would also like to thank Dr. Valentin Iota (Lawrence Livermore National Lab., California) for valuable comments and suggestions.


\end{document}